\documentclass[compsoc, conference, a4paper, 10pt, times]{IEEEtran}

\usepackage{cite}
\usepackage{amsmath,amssymb,amsfonts}
\usepackage{algorithmic}
\usepackage{graphicx}
\usepackage{textcomp}
\usepackage{xcolor}
\usepackage{booktabs}
\usepackage[hidelinks]{hyperref}

\begin{document}

\title{Towards a New Configurable and Practical \\Remote Automotive Security Testing Platform
}

\author{\IEEEauthorblockN{Sekar Kulandaivel}
\IEEEauthorblockA{Robert Bosch Research \& Technology Center\\
sekar.kulandaivel@bosch.com}
\and
\IEEEauthorblockN{Wenjuan Lu}
\IEEEauthorblockA{Block Harbor Cybersecurity\\
wen@blockharbor.io}
\and
\IEEEauthorblockN{Brandon Barry}
\IEEEauthorblockA{Block Harbor Cybersecurity\\
brandon@blockharbor.io}
\and
\IEEEauthorblockN{Jorge Guajardo}
\IEEEauthorblockA{Robert Bosch Research \& Technology Center\\
jorge.guajardomerchan@bosch.com}}
\author{
\IEEEauthorblockN{
Sekar Kulandaivel,\IEEEauthorrefmark{1}
Wenjuan Lu,\IEEEauthorrefmark{2}
Brandon Barry,\IEEEauthorrefmark{2} and
Jorge Guajardo\IEEEauthorrefmark{1}}
\IEEEauthorblockA{
\IEEEauthorrefmark{1}\textit{Robert Bosch LLC --- Research and Technology Center},
\IEEEauthorrefmark{2}\textit{Block Harbor Cybersecurity}\\
\{sekar.kulandaivel, jorge.guajardomerchan\}@bosch.com,
\{wen, brandon\}@blockharbor.io
}}

\maketitle

\begin{abstract}
In the automotive security sector, the absence of a testing platform that is configurable, practical, and user-friendly presents considerable challenges. These difficulties are compounded by the intricate design of vehicle systems, the rapid evolution of attack vectors, and the absence of standardized testing methodologies. We propose a next-generation testing platform that addresses several challenges in vehicle cybersecurity testing and research domains. In this paper, we detail how the Vehicle Security Engineering Cloud (VSEC) Test platform enables easier access to test beds for efficient vehicle cybersecurity testing and advanced (e.g., penetration, fuzz) testing and how we extend such test beds to benefit automotive security research. We highlight methodology on how to use this platform for a variety of users and use cases with real implemented examples.
\end{abstract}

\section{Introduction}

In light of recent security standards~\cite{international2021iso} and regulations~\cite{unece2021regulation}, the automotive cybersecurity industry demands more efficient and widespread cybersecurity testing of automotive products. For example, ISO/SAE 21434 requires verification and validation (V\&V) testing among the vehicle, its subsystems, and its components. Types of testing include cybersecurity functional testing, vulnerability scanning, penetration testing, and fuzz testing. This cybersecurity testing is further complicated by the current push towards software-defined vehicles, which weaves safety with security and demands faster cycles of development (and thus equally faster cycles of testing).

Unfortunately, executing security tests poses a substantial challenge as a result of limited access to hardware, the lack of configurable networks, the difficulty of connecting and configuring automotive electronic control units (ECUs) for non-specialists, and the lack of qualified personnel. Fully-enabled cybersecurity features often limit the capability of functional test engineers to be able to measure and calibrate features on the spot, which results in these critical features being disabled for the majority of the test cycle. Cybersecurity testing is usually performed after functional and environmental testing, which can make it challenging for security teams to push for patching identified vulnerabilities if significant changes are required~\cite{on:vicone-report}.

In addition to the lack of thorough security feature testing, penetration testing on complete systems is also an area that needs much improvement. It is already difficult to find security experts who are well-versed in vehicle design and infrastructure, and the physical nature of the vehicle means often these experts must be hands-on with the entire vehicle to find the path of least resistance for a threat. The size and price tag of the targets makes penetration testing of vehicles a costly process. Likewise, the ECUs within a vehicle are complicated to set up outside of a running vehicle, especially without access to proprietary knowledge (e.g., software, wiring diagrams).

There are existing efforts to combat these challenges and make automotive security testing easier. Some solutions focus on cloud-based tests with virtual ECUs~\cite{cloudcar}; however, these are typically isolated and may not represent a whole system. Another solution is hardware-in-the-loop (HIL) testing~\cite{vectorHIL, on:keysight}. This type of testing approach is often tailored for testing individual ECUs or (at most) a couple of ECUs, and it would be difficult and costly to replicate a whole vehicle network of physical ECUs as it would require several HIL systems. In addition, custom test beds are often manually built to target a specific function, but these test beds are not configurable enough to be used efficiently and cost-effectively for all necessary security testing. Due to this effort, instrumenting these test beds and connecting them to a larger testing infrastructure has not traditionally been a common approach.

There has been some significant progress in automotive security testing. The Eclipse openDuT framework~\cite{opendut} aims to automate testing and validation for automotive software and applications by offering a description language for test cases, management of test cases, and orchestration of distributed ECUs. VitroBench~\cite{vitrobench} connects ECUs on a vehicle network bus to an FPGA that can launch a variety of network attacks while simulating signals from vehicle sensors. CANlay~\cite{jepson2023canlay} offers a network abstraction layer to facilitate transmission of controller area network (CAN) data and sensor signals between (remote) ECUs. Additionally, CRATE~\cite{bgCRATE} offers an all-in-one-box hardware solution that simplifies the connection to multiple ECUs and vehicle networks and offers other convenient capabilities, such as dedicated power switching. While each of these works targets some aspect of automotive security testing, a platform that combines many of these features into a single configurable, user-friendly, and practical remote security testing platform still eludes the industry.

In this paper, we build such a testing platform and explore several testing methodologies that this platform enables for penetration testers, ECU engineers, and security researchers. We share our design and vision for this testing platform, and we posit that such a platform should offer the following key characteristics: (1) managed multiple user access control in a familiar environment, (2) access to real hardware with a realistic, configurable network, (3) hardware support at centralized locations with remote user access, and (4) ability to add custom applications for security testing.

\section{Current Security Testing Platforms}

Currently, we see two main categories of verification activities for vehicle cybersecurity that testing platforms have aimed to support. As a part of automotive engineering development activities, cybersecurity verification is done as a part of a standard V\&V process. The usual procedure for aligning this with ISO/SAE 21434 Clause 10.4.2 is as follows:
\begin{enumerate}
    \item \textit{Verification methods:}
    This type of testing usually follow the ASPICE engineering process~\cite{on:aspice} of Unit Testing, Integration Testing, and Qualification Testing. Cybersecurity requirements are combined with functional requirements and verified as per engineering process standards. This type of testing includes requirement-based testing, interface testing, resource evaluation, control and data flow verification, and static and dynamic analysis. The techniques used in this type of testing follow general software testing methodology and can be easily performed by a qualified test engineer.
    \item \textit{Testing for unidentified weakness or vulnerabilities:}
    This type of testing is unique to the field of cybersecurity and requires the test engineer to have specialized knowledge of or experience with cyber attacks (e.g., the Sam Curry attacks~\cite{on:samcurryhack2023}) in order to be effective. These activities include targeted functional testing, vulnerability assessment, fuzz testing, and penetration testing.
\end{enumerate}

Out of the numerous cybersecurity testing activities discussed above, the verification testing methods are mostly addressed with a comprehensive engineering process. These methods stand to benefit from broader acceptance of other automotive quality-oriented disciplines, such as ASPICE and functional safety, which should lead to improvements. As the industry adopts better and more effective software quality engineering, in part, thanks to regulatory pressure, we are starting to see many tools being developed to address the activities of vulnerability assessment and fuzz testing, specifically targeting automotive use cases.

\paragraph{\textbf{Challenges}}
Even with the entrance of more service providers into the field, automotive cybersecurity functional testing and penetration testing, particularly for complex components or systems, still suffer from shortcomings in technical proficiency and cost effectiveness. Below, we list some of the main challenges we see in these types of testing:
\begin{enumerate}
    \item Technically qualified automotive cybersecurity testing specialists are hard to find,
    \item Testing engagements are difficult to manage due to the (implicit) need to access hardware remotely, and
    \item Fully remote penetration testing is still not realistic for most automotive use cases and difficult to repeat over time.
\end{enumerate}

As evident in other cybersecurity disciplines, such as web security or Internet-of-Things (IoT) security, a test engineer should readily have access to targets. However, getting access to a test target is currently much more difficult for automotive pentesters due to the cost, availability, complexity of modern vehicle architectures, and equipment needed for car hacking. Due to the unique skill set of pentesters and the small talent pool, most automotive OEMs and suppliers do not have the internal capability to perform fuzzing or penetration testing on every product. Penetration testing requires a completely different skill set and mindset from the traditional test engineer or development engineer. This activity is increasingly outsourced to third-party service providers. Compared to internal V\&V, this outsourcing is much more difficult to manage in the event of project delays, dependency on third-party expertise and resources, concerns about data security and confidentiality, and limited hardware availability, which can result in timeline penalties and potentially cause mistakes in test target configurations.

There has been some effort in addressing these issues with running software simulations of ECUs and networks on the cloud. However, these are often extremely limited and do not represent even a fraction of all attack surfaces of a modern vehicle~\cite{mahmood2021automotive}. More importantly, any hardware-based weaknesses and vulnerabilities are often completely absent in software-only simulations. Traditionally, test beds for security projects are custom built for penetration testing with a highly customized interface. These test beds have a short lifetime (i.e., the length of a penetration testing contract), are very time consuming to build, and require specialized knowledge to interface with them. Due to the custom nature of these test beds and their fast turnarounds, it has not been worthwhile to build a generalized testing platform. As vehicles evolve to become more software-defined, a high demand is placed on testing far beyond product development and into the production and operation stages of the vehicle life cycle. Without a test bed to rerun penetration tests on, identifying new vulnerabilities during the life of a product can be difficult.

\subsection{Recent progress}

\begin{table*}[ht]
    \centering
    \begin{tabular}{ccccccc}
    & \multicolumn{6}{c}{\textbf{TESTING PLATFORMS}} \\
    \textbf{KEY ATTRIBUTES} & \cite{opendut} & \cite{bgCRATE} & \cite{vitrobench} & \cite{cloudcar} & \cite{jepson2023canlay} & This platform \\
    \midrule
    Remote access & \checkmark & \checkmark &  & \checkmark & \checkmark & \checkmark \\ \midrule
    Rich web interface &  &  &  & \checkmark &  & \checkmark \\ \midrule
    Physically distributed ECUs & \checkmark &  &  &  & \checkmark & \checkmark \\ \midrule
    Hands-free configurable ECU network &  & \checkmark & \checkmark &  & \checkmark & \checkmark \\ \midrule
    Handles physical COTS ECUs & \checkmark & \checkmark & \checkmark &  & \checkmark & \checkmark \\ \midrule
    \end{tabular}
    \caption{A comparison of key attributes in our testing platform versus other platforms and frameworks}
    \label{tab:comparison}
\end{table*}

Despite the challenges faced by the industry, we have seen considerable effort being made recently to address the shortcomings of cybersecurity testing. Table~\ref{tab:comparison} compares our platform against several other modern testing platforms and highlights the key attributes that we offer and also explore in this paper.

\paragraph{\textbf{Testing hardware}}
Electronics test and measurement equipment and software manufacturer, Keysight Technologies, launched the SA8710A Automotive Cybersecurity Test Platform~\cite{on:keysight} a few years back with hardware that connects to all relevant interfaces on a modern vehicle and built-in security scanners and fuzzing tools. Built with its product line of connectivity and measurement instruments, the Keysight platform offers hardware that is capable of capturing small variations and disturbances in communication signals. Last year, we saw the launch of the BG Networks CRATE~\cite{bgCRATE}, an all-in-one hardware box capable of connecting to most of the vehicle interfaces, with options to expand using USB. The CRATE™ also offers some software tools to facilitate remote access and fuzz testing. Other traditional hardware test interface providers do not offer specific hardware for security testing, but they offer software plug-ins to support limited security testing~\cite{vectorHIL}.

\paragraph{\textbf{Testing frameworks}}
The Eclipse openDuT~\cite{opendut} framework focuses on automated and repeatable test and validation executions in the automotive space. This work is still in an incubating state and eventually aims to support a number of use cases, such as fully automated gray-box tests, tests across distributed test benches, and other functional tests. This framework provides end-to-end encryption of a private network between edge devices (ECUs, Restbus simulations, etc.) with an underlying control and registration logic. Ideally, network traffic from different devices under test can be linked together to create a more unified approach for testing. Additionally, CANlay offers virtual configurable networks where isolated ECUs can be bridged with sensor simulators for J1939 testing. A subset of ECUs here called overlays can be tested by carrying network traffic between these isolated ECUs.

\paragraph{\textbf{The ideal solution}}
In contrast to prior work, we aim to build a platform that is easy to use by many users without deep hardware knowledge and preferably provide remote access to enable global teams to work on the same project at the same time. The platform should be capable of offering multiple test beds and, if required, the test bed should be configurable so that only a subset of ECUs can be operated on-demand. This hardware should be located in a central location with staff on site that specialize in hardware to support remote users that access the system. It should also be easy for a remote test engineer to launch their own custom security tests using remotely-programmable hardware and measurement tools.

\section{A New Testing Platform}

We present a flexible and configurable testing platform for automotive security by extending a cloud-based framework with additional components to support a number of testing methodologies. We focus on building a solution that is useful for many types of users, ranging from penetration testers to security researchers, and we aim to create a platform that bridges features in both software and hardware testing. We do not aim to cover all aspects of functional and safety testing, and we do not attempt to recreate a HIL system. We implement some of these testing methodologies and discuss practical benefits of our tests.

\subsection{Testing platform roles}

We design this testing platform with a variety of users and roles in mind:\\
\textit{Penetration testers:} Users who need to test quickly and easily launch automated pentesting scripts, perform wireless tests while remote, access platforms without needing to ship hardware, connect a local device to a remote device (e.g., OEM dealership tool), and easily switch between different test beds from different contracts.\\
\textit{Researchers:} Users who need to easily deploy and demonstrate novel security research on ECUs from multiple OEMs, copies of the same ECU for repeatability, and multiple model years while performing analysis with measurement tools (e.g., oscilloscopes, logic analyzers) and launching attacks from a custom ECU on the network.\\
\textit{Engineers:} Users who needs managed access to the platform via scheduling, global remote access to test beds, quick reconfiguration of a prior project to handle new vulnerabilities, and a familiar user interface to interact with.\\
\textit{Lab technicians:} A role that is responsible for setting up hardware in the lab, activating hardware-only features in-person, and moving probes for measurement tools if need.\\

\subsection{Web-based remote access}

The first step to developing a user-friendly testing platform is providing an easy-to-use and familiar interface. Vehicle Security Engineering Cloud~\cite{blockharborVSEC} (VSEC) is a cloud-based platform that aims to connect and automate various steps of the vehicle cybersecurity engineering process. A major problem we see in the V\&V processes is the lack of repeatability and frequency of tests. To address this problem, while maintaining confidentiality, VSEC Test centralizes the testing process and execution while distributing the efforts across locations and resources.
 
\paragraph{\textbf{Managed remote access}} Via a web interface, the VSEC Test platform allows users to register and connect hardware test benches or vehicles to the cloud, which are then shared and managed centrally under an enterprise account. Authorized users are able to schedule benches, access all available interfaces of the bench via a Linux terminal, and control multiple power sources to individual configured targets. User access is configured and controlled by the lab orchestration software to prevent unauthorized access. Features such as file explorer and workspace sharing can be enabled to allow quick ways to reproduce test environments; otherwise, each user has a sandbox environment in a Docker container to keep their workspace separated. In addition, VSEC Test also hosts a test management system that allows for continuous and repeated test setup with connected benches.

To evaluate the effectiveness and robustness of the VSEC Test platform, we set up a few bench and vehicle targets in the lab and connected various vehicle interfaces to a local machine that connects to VSEC Test on the cloud. We used built-in tests to remotely scan for basic information from a complete vehicle network, and compared the results with similar open-source tools installed directly to the local machine. Then we set up a custom test case on VSEC Test to be ran periodically on the remote target while purposely triggering some network downtime during test execution to simulate unplanned network disruptions, and compared the results to a local setup. We found that the added latency due to the cloud connection was not sufficient to cause a noticeable negative effect on the effectiveness of the test team. Test results obtained from the cloud platform are functionally equivalent to those achieved through a local setup. Moreover, the cloud platform provides the added benefit of facilitating instant sharing of results among team members and automating the report generation process, thereby eliminating the need for manual compilation of results from open-source tools.

\subsection{Testbed setup}

\paragraph{\textbf{Sourcing ECUs}}
In the context of this work, we define ECUs as production ECUs from the field (i.e., real vehicles), engineering samples (e.g., A-samples, B-samples), or even off-the-shelf automotive-grade microcontrollers (MCUs). Depending on the user, obtaining ECUs for the testing platform will require different approaches. Where an engineer and pentester would get hardware from their own supply chain or from the contracted customer, respectively, a researcher would likely obtain ECUs from a real vehicle from either a salvage yard or buying ECUs directly from OEM parts shops. Testing a single ECU can be sufficient if the tests are focused on purely CAN physical layer specifics or other single-ECU session-based protocols, such as Unified Diagnostic Services (UDS) protocol and Universal Measurement and Calibration Protocol (XCP). In some cases, multiple ECUs are needed to wake up certain ECUs (e.g., a powertrain ECU may require a body ECU and vice versa), although we have reverse-engineered some of these wake-up messages. For our platform, we source ECUs across three different OEMs from a local salvage yard and from a real vehicle: five powertrain ECUs (two pairs of duplicates from different model years), a gateway ECU, a power steering ECU, and two instrument panel clusters (IPCs). We also acquire a C-sample gateway ECU from one of the aforementioned OEMs, which matches the OEM of one of the IPCs. Additionally, we also obtain a CAN-enabled automotive-grade MCU with full programming capabilities.

\paragraph{\textbf{Configurable network}}
Most users will often work on projects that require ECUs with different bus speeds, different OEMs, and different attached hardware (e.g., measurement tools, actuators, etc.) As such, it is necessary for a testing platform to provide the ability to configure the platform remotely. The alternative here is having access to multiple real vehicles or (re)building separate test beds, which is expensive and harder to manage. We build a prototype of a configurable CAN bus network system that supports multiple bus speeds, multiple ECUs from multiple OEMs, and the ability to switch on/off the power of each ECU on-demand via VSEC Test.

\begin{figure*}
    \centering
    \includegraphics[width=0.8\textwidth]{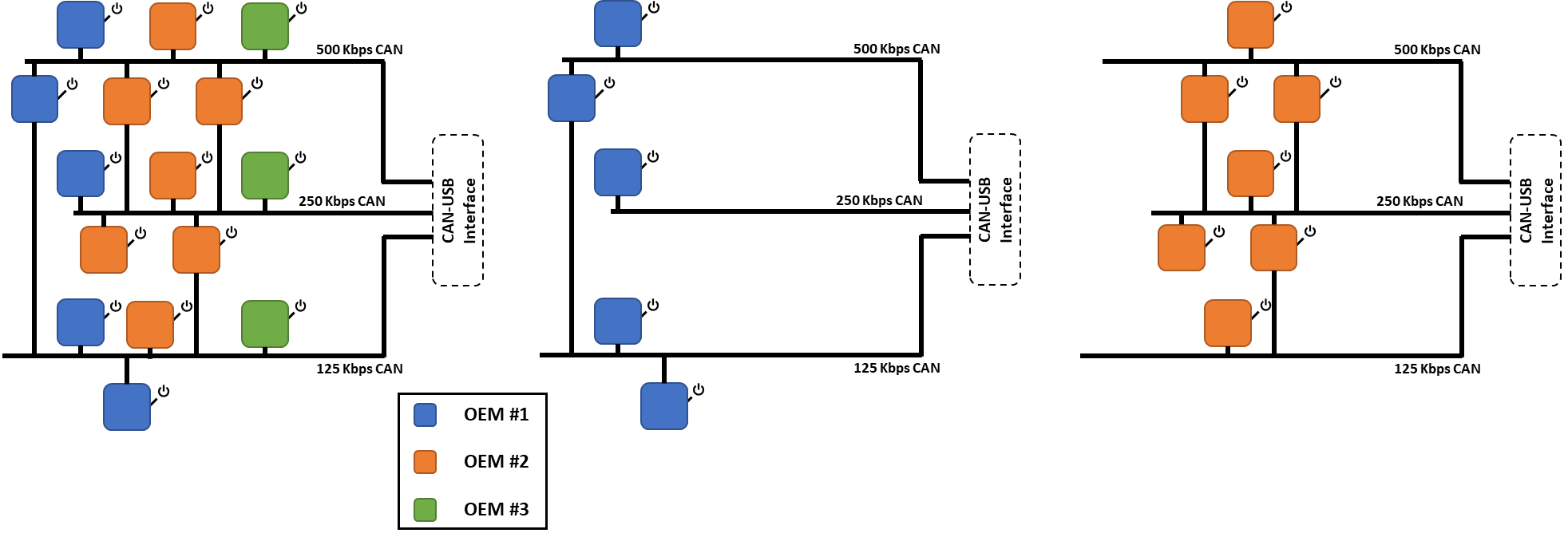}
    \caption{
    The entire physical network layout (left) can contain ECUs from a number of different original equipment manufacturers (OEMs). These ECUs may connect to CAN buses with different speeds, which ultimately connect to a PC via a CAN-USB interface. By powering off all other ECUs except those from OEM \#1, we can construct a network consisting of ECUs from just OEM \#1 (middle). Likewise, we can construct a network of ECUs from just OEM \#2 (right) by powering off all other ECUs.}
    \label{configurable-network}
\end{figure*}

In Figure~\ref{configurable-network}, we depict an example of of a single platform that: (1) hosts multiple buses with different bus speeds, (2) multiple OEM platforms on one bench, and (3) the ability to control the network topology. The figure demonstrates what the entire bench looks like with ECUs from three different OEMs as well as how our testing platform would bring up a network consisting of just ECUs from OEM \#1 and OEM \#2. Using a relay board that powers on/off each ECU on command, a tester could also isolate a single ECU or subset of ECUs from the entire set. The ability to slowly ramp up the number of ECUs under test is useful when developing a technique (e.g., a new attack), and the ability to test a technique against multiple OEM implementations can determine the applicability of the technique. For our platform, we construct a configurable network by attaching the power supply of each ECU to a 16-port relay board,\footnote{The ignition pin can be powered by a separate relay if needed, thus having up to 16 ports can help control both main power and ignition while maximizing the number of ECUs under test.} which is controlled by an Arduino Due MCU. A script on the tester PC communicates with this MCU over Serial, and the code on the Arduino toggles the user-specified ECU to activate/deactivate. In this work, we focus on CAN buses, and adding a new bus protocol such as Automotive Ethernet or CAN-FD would be similar.

\paragraph{\textbf{Network simulation}}
Real CAN traffic from road tests is often only available after a vehicle is in production. This traffic is usually the only information a researcher can access to simulate the bus network. Reverse-engineering CAN traffic can successfully uncover which CAN messages control which vehicle features/sensors~\cite{pese2019librecan, vitrobench}. Engineers and some pentesters (depending on the contract) have access to proprietary CAN database (DBC) files, which identify the producer/consumer of each signal for all CAN messages. An alternative source of information for engineers and pentesters are Restbus simulations, which provides sufficient CAN traffic for the ECU under test to function as expected. For our platform, we investigated another method of obtaining a larger network of ECUs by combining the traffic from multiple ECUs that belong to the same make and model but are physically-distributed. We implemented a test bed with an IPC located in Pennsylvania and remotely connected to a powertrain ECU hosted in Michigan. While there is some latency involved with forwarding CAN traffic between these two locations, the latency is small enough to enable access to most UDS commands and exhibit control of the IPC without any obvious delays.

\paragraph{\textbf{Environment simulation}}
While reverse-engineering and proprietary DBC files can enable users to interpret CAN signals, it is still challenging to mimic traffic from road tests without spending hours driving a vehicle. Users can utilize driving simulators, such as CARLA~\cite{on:carla}, to provide hours and miles of simulated road traffic. Combining these simulators with either a reverse-engineered or original DBC file can generate realistic CAN bus traffic to the ECUs under test. Engineers often have access to advanced sensor and actuator models that can significantly enhance the realism of the CAN bus traffic. While these models are typically deployed in software-only simulations, they could be connected to our testing platform to capture results only observed on real hardware. For our testing platform, we utilize CARLA in combination with a DBC file to provide realistic speed values to an IPC and then deploy an intrusion detection system (IDS) from prior work~\cite{gehrer2022system} as if it were in a real car.

\paragraph{\textbf{Measurement tools}}
We expand the capabilities of the platform by connecting measurement tools, such as oscilloscopes and logic analyzers, and offering software control of these tools. As these tools are often expensive, a platform with centralized hardware can reduce overall hardware costs by permitting shared access to these tools. For our platform, we connect a Picoscope oscilloscope to the power pins of an ECU and a Saleae logic analyzer to all of our CAN buses. Via command-line scripts, the Picoscope can capture power profiles as input to the aforementioned IDS. Additionally, the logic analyzer can observe bit-level CAN bus traffic for the purpose of building advanced CAN bus attacks~\cite{kulandaivel2019canvas, kulandaivel2021cannon, kulandaivel2021revisiting}.

\section{Enabled Testing Methodologies}

\subsection{Secure Development Lifecycle (SDL) testing}

VSEC Test aims to integrate all methods of vehicle cybersecurity testing into a single easy-to-use platform. As discussed above, a portion of cybersecurity verification, as required by the ISO/SAE standard, is no different than software and systems verification required by other industry standards. The remote capabilities and test management features offered by VSEC Test allows test engineers to design and set up the test procedures one time in order to run on a schedule or triggered by other tools as a part of CI/CD frameworks. This centralized platform for test procedures and results can potentially be used to exchange data throughout the supply chain and ensure consistent verification test coverage for the development life cycle of a vehicle.

For example, we perform continuous cybersecurity functional testing on components with over-the-air (OTA) update functionality as a service. We set up the test rack and develop the test procedures once on VSEC Test, and it is performed on a periodic schedule and also whenever a new software update is applied. Continuous functional testing against cybersecurity requirements effectively identified vulnerabilities that were overlooked during the development cycle. In addition, VSEC Test provides a set of built-in tests as a preliminary scan for known weaknesses and vulnerabilities, which can be used directly against a target vehicle or ECU. This scan can allow users to easily and quickly determine if there are large gaps in measures protecting the interface or if there are known exploits that will work on the target vehicle or ECU.

\subsection{Penetration testing}

\paragraph{\textbf{Pentesting approaches}}
Most pentest engagements start with an exploratory phase, where the tester aims to discover as much information about the target as possible. The duration and effort in this phase is largely determined by the amount of information available to the tester. VSEC Test allows users to build a pool of their own test scripts or use built-in discovery scans specifically targeting automotive networks, including protocols such as UDS and XCP. These tests can be queued up and ran in the background or over the weekend, saving time and allowing a team to focus on more technically challenging tasks or other projects. The web interface allows testers to access and check on test status at any time from anywhere to determine the next step of action.

\paragraph{\textbf{Partner Pentesting}}
A major difficulty in accessing the talent pool of world-class vehicle penetration testers is physical location. The VSEC Test platform enables \textit{any} remote engineer with credentials to connect to the bench and handle any physical interactions with the component and remotely script and execute tests from VSEC Test. We call this Partner Pentesting.

\begin{figure*}
    \centering
    \includegraphics[width=0.75\textwidth]{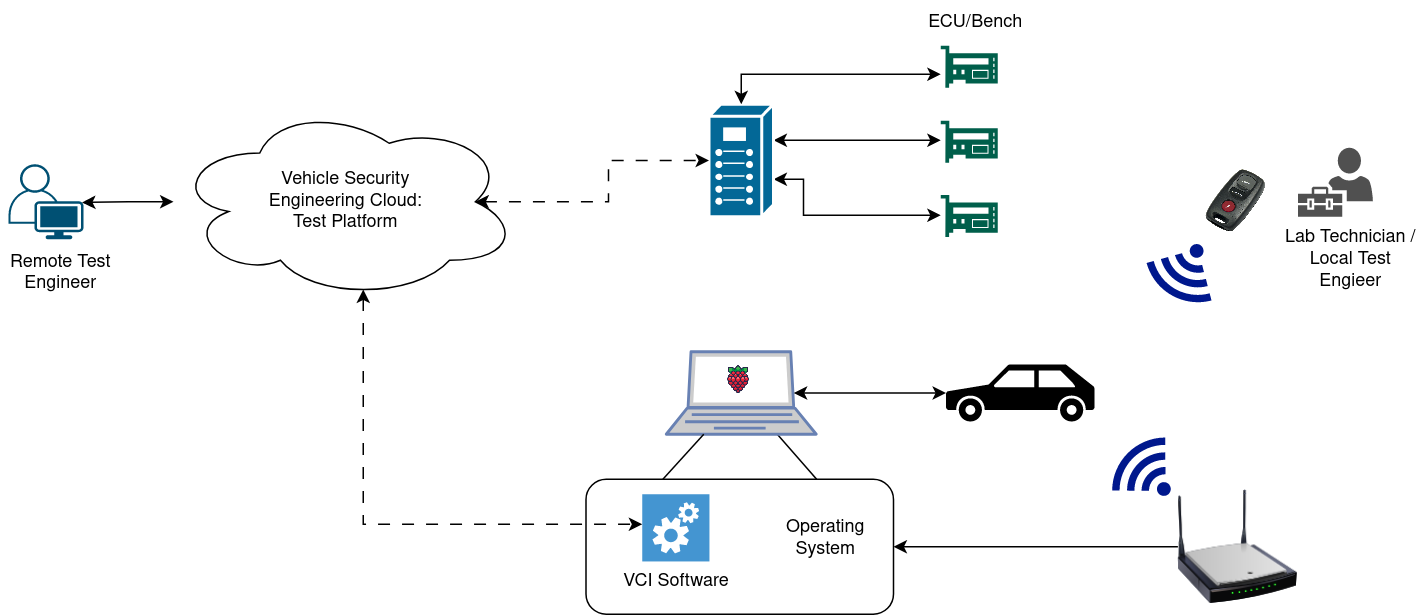}
    \caption{The remote test engineer can access either a fully remote testing platform with access to multiple test beds or work collaboratively with an in-lab (i.e., local to the lab) test engineer for tests that require hands-on support. This Partner Pentesting method enables the technical specialist to focus on their testing strategies while leaving the hardware setup and physical controls to the lab technician/local test engineer.}
    \label{remote-pentesting}
\end{figure*}

To evaluate the Partner Pentesting concept, we set up a scenario as depicted in Figure~\ref{remote-pentesting}, where a \textit{remote} wireless penetration testing specialist is working with a local test engineer to control a software-defined radio connected to the cloud and capture communication between the key fob and the car in addition to any vehicle network messages. The specialist then proceeds to study and decode the traffic, create scripts to attack the communication, and work together with the local test engineer to validate findings. In this case, without physical access to the vehicle and a local partner to perform tasks, it would have been very difficult to perform a meaningful penetration test against the vehicle's wireless entry system.

\paragraph{\textbf{Side-channel analysis}}
On the other hand, we have other benches in the lab can be tested with little to no user interactions. For these targets, fully remote network-level penetration test with a test bench setup can achieve results on-par with on-site engagements. However, we found that more hands-on penetration testing procedures such as hardware analysis and side channel attacks are still very difficult to perform without having an onsite specialist. As a result, while acceptable for many scenarios, fully remote penetration testing remains very limited for certain configurations due to manual activation required for physical interfaces and required hardware and tooling interaction for certain procedures. However, we are able to make progress with a Partner Pentesting setup that provides assistance from onsite personnel and achieves comparable outcomes as a fully onsite engagement.

\subsection{Research testing}

We now discuss how our testing platform can enable a researcher to rapidly establish and experiment with numerous ECU networks to support their security research. A testing platform where the ECUs present on the bus can be easily configured to support the development of a research project is ideal for fine-tuning an attack or defense.

\paragraph{\textbf{Enhancing test bed functionality}}
For research use cases, it is critical to have a programmable ECU on the network to launch attacks or implement defenses. Typically, simulation-based environments remove real-world characteristics, such as bus voltage, bit-level CAN bus arbitration, bus errors, etc. Likewise, advanced attacks beyond simple CAN bus injection often require microcontroller-level timing precision that cannot be achieved by the PC-USB interface. Thus, our platform includes an automotive-grade MCU to mimic having programming access to an ECU. For other testing purposes, a bench with MCUs from several vendors could enable a user to test with different vendor libraries and MCU features very quickly.

\paragraph{\textbf{Exploring real research use cases}}
We identify a set of research projects that use a similar test setup and demonstrate how our testing platform could have made that research easier and more efficient. To demonstrate the usefulness of this new remote and configurable platform, we implement and test three different open-source research implementations using a single bench of ECUs. We remotely run the CANvas network mapper~\cite{kulandaivel2019canvas} to correctly identify unique ECUs on two CAN buses with different speeds and across several configurations network topologies. We also remotely demonstrate the CANnon bus disruption attack~\cite{kulandaivel2021cannon} using the automotive-grade microcontroller on our bench and observe numerous faults on the CAN bus, which were captured by the attached Saleae logic analyzer. Finally, we remotely deploy techniques from the CANdid authentication bypass attack~\cite{kulandaivel2021revisiting} against three configurations of a single bench, where we isolate three powertrain ECUs from three different vehicle models using our relay-controlled power inputs and observe the same ability to control the randomness of the challenge. The CANnon and CANdid attack require launching attacks from an automotive-grade microcontroller and require an attached oscilloscope or logic analyzer to fine-tune certain parameters, which are all features offered by our testing platform.

\paragraph{\textbf{Offering new services}}
A unique challenge with automotive security research is the lack of hands-on demonstrators for the community to experiment and test with. We envision the offering of new services that permit access to research examples of attacks, defenses, measurement techniques, etc. Instead of limiting the availability of research to published code and papers, the ability to remotely access a running sample in a real in-vehicle network environment would strengthen the research community. In cases where code or hardware descriptions do not want to be publicly disclosed, our testing platform could enable such access.

\section{Conclusion \& Outlook}

With vehicles becoming more software-defined, the need for higher quality and more automated security testing is evident. A flexible, easy-to-use, and remote-capable test platform not only improves the current industry testing capabilities, but it can also potentially provide easier access to logistically challenging and costly hardware. This allows the industry access a larger pool of talent for testing and allows more people to learn about automotive components. Furthermore, we see potential in leveraging this type of cloud-based platform to reduce costs and increase participation for global hacking competitions and Bug Bounty programs.

That being said, the ideal testing platform is one that the community will actually use and implement in their research, development, and production. In this work, we propose several well-defined methodologies to build a useful and configurable remote testing platform. However, as the needs of the automotive security industry changes, this testing platform should be able to adapt and be as flexible as possible. Once other testing platforms mature, future work should investigate the different underlying technologies and features used to build these platforms and explore relevant trade-offs and limitations.

\bibliographystyle{plain}
\bibliography{references}

\end{document}